\documentclass[%
 prl,
 amsmath,amssymb,
 reprint,%
]{revtex4-1}

\usepackage{graphicx}% Include figure files
\usepackage{dcolumn}% Align table columns on decimal point
\usepackage{bm}% bold math
\begin{document}

\title{Violation of Bell Inequalities with a Mixture of Separable States\\in a Multiple-Photon Absorption Attack on Ekert Protocol}%
%\title{Violation of Bell Inequalities with Separable States\\in a Multiple-Photon Absorption Attack on Ekert Protocol}%
%\title{Violation of Bell Inequalities with a Mixture of Separable States}%
%\title{Multiple-Photon Absorption Attack on Ekert Protocol}%

\author{Guillaume Adenier}
\email{adenier@rs.noda.tus.ac.jp}
\author{Noboru Watanabe}%
\affiliation{Tokyo University of Science, 2641 Yamazaki, Noda, Chiba 278-8510, Japan}%
\author{Andrei Yu. Khrennikov}
\author{Irina Basieva}
\affiliation{Linnaeus University, Vejdes plats 7, SE-351 95 V\"{a}xj\"{o}, Sweden}%

\begin{abstract}
We propose a new type of attack on Ekert protocol in which, rather surprisingly, Eve drives a violation of Bell inequalities in Alice's and Bob's detectors with a mixture of separable states. She does so by sending correlated pulses containing several photons at frequencies where only multiple-photon absorptions are possible in their detectors. Whenever the photons stemming from a same pulse are dispatched in such a way that the number of photons is insufficient to trigger a multiple-photon absorption in either channel, the pulse remains undetected. We show that this simple feature leads to violations of Bell inequalities that can match closely those predicted for entangled states, even in the simplest cases of two-photon and three-photon absorptions.
\end{abstract}

\maketitle
Ekert protocol \cite{Ekert91} uses entangled states to guarantee the secrecy of a key distributed to two parties (Alice and Bob) desiring to communicate secretly through a public channel. Identical measurements performed on a maximally entangled state yield a perfect correlation \cite{EPR35} that can be used to produce a shared key, while its secrecy is guaranteed by a violation of Bell inequalities \cite{Bell64,CHSH,Aspect82,Weihs98,Khrennikov} measured for non-identical measurements.

We propose a new type of attack on this protocol, in which Eve controls a source of separable states with which she tries to convince Alice and Bob that the key that they extract with it is guaranteed by the observation of a violation of Bell inequalities, while in fact Eve has a full knowledge of the local states sent to them both.

\section{Single-photon absorption attack}
In this first attempt, Eve is trying to approach as best as she can the predictions of Quantum Mechanics for a singlet state, but using only a mixture of separable states, in a single-photon absorption scenario. She will not succeed in getting a violation of Bell inequalities, but this result will serve as a basis to obtain a significantly better result in the case of multiple-photon absorption \cite{Braunstein61}.

Eve prepares an ensemble of pairs of pulses for Alice and Bob. Each pulse contains exactly one photon that is linearly polarized in a direction $\lambda_i$ chosen by Eve. It can thus be described in its own Hilbert subspace $\mathcal{H}_i$ by a two-level state of the form
\begin{equation}\label{E:10}
    |\lambda_i\rangle=\cos\lambda_i|0\rangle+\sin\lambda_i|1\rangle,
\end{equation}
where $|0\rangle$ and $|1\rangle$ are the eigenvectors of $\hat{\sigma}_\mathrm{z}$ with eigenvalues $-1$ and $+1$ respectively.
 The pair of pulses can then be described in the tensor product Hilbert space $\mathcal{H}_{12}=\mathcal{H}_1\otimes\mathcal{H}_2$ as $|\Lambda_{12}\rangle=|\lambda_1\rangle\otimes|\lambda_2\rangle$.

Alice and Bob, who are performing local measurements respectively in $\mathcal{H}_1$ and $\mathcal{H}_2$, want to measure the statistical correlation of the pairs they receive from Eve when they perform some rotations $\hat{R}(\theta_\mathrm{A})$ and $\hat{R}(\theta_\mathrm{B})$ on their respective pulses, followed by a measurement of the observable $\hat{\sigma}_\mathrm{z}$.

Since Eve wants them to obtain an as good (anti)correlation as possible with separable states, she sends orthogonal states to Alice and Bob, that is, such that $\lambda_1=\lambda$ and $\lambda_2=\lambda+\frac{\pi}{2}$. For a pair initially represented by $|\lambda\rangle\otimes|\lambda+\frac{\pi}{2}\rangle$, the state after local rotations on each sides becomes
\begin{equation}\label{lambdaAB}
    |\Lambda_\mathrm{AB}\rangle=|\lambda+\theta_\mathrm{A}\rangle\otimes|\lambda+\frac{\pi}{2}+\theta_\mathrm{B}\rangle,
\end{equation}
which can be expanded as
\begin{equation}\label{A-5}
\begin{aligned}
    |\Lambda_\mathrm{AB}\rangle=\cos(\lambda+\theta_\mathrm{A})\cos(\lambda+\frac{\pi}{2}+\theta_\mathrm{B})&|00\rangle\\
    +\sin(\lambda+\theta_\mathrm{A})\cos(\lambda+\frac{\pi}{2}+\theta_\mathrm{B})&|10\rangle\\
    +\cos(\lambda+\theta_\mathrm{A})\sin(\lambda+\frac{\pi}{2}+\theta_\mathrm{B})&|01\rangle\\
    +\sin(\lambda+\theta_\mathrm{A})\sin(\lambda+\frac{\pi}{2}+\theta_\mathrm{B})&|11\rangle,
\end{aligned}
\end{equation}

Keeping the condition of orthogonality between Alice and Bob from pair to pair, Eve is randomizing the parameter $\lambda$ associated to each pair. The state of the ensemble of pairs prepared by Eve can therefore be described by a mixture. We can characterize the probability to obtain a state lying between $|\lambda\rangle$ and $|\lambda+d\lambda\rangle$ by a probability density distribution $\rho(\lambda)$ on a single probability space $(\Omega,\mathcal{F},P)$.

The mixture $\hat{\rho}$ describing the ensemble of pairs after rotation is therefore
\begin{equation}\label{mixture}
    \hat{\rho}_\mathrm{AB}=\int_\Omega\rho(\lambda)d\lambda \:|\Lambda_\mathrm{AB}\rangle\langle\Lambda_\mathrm{AB}|,
\end{equation}
with $\int_\Omega\rho(\lambda)d\lambda =1.$

Note that we chose to denote the polarization of the pulses prepared by Eve as $\lambda$, a symbol that is traditionally reserved to hidden-variables, but this variable is in fact hidden only from Alice and Bob, and it is not an hypothetical supplementary parameters to Quantum Theory.

After performing their respective rotation on the received pulses, Alice and Bob perform a joint measurement $\sigma_\mathrm{z}\otimes\sigma_\mathrm{z}$ on each pair. The expectation value of this measurement is:
\begin{equation}\label{A-7}
    E_\mathrm{AB}=\langle\hat{\sigma}_\mathrm{z}\otimes\hat{\sigma}_\mathrm{z}\rangle_{\hat{\rho}_\mathrm{AB}}=\textrm{Tr}(\hat{\rho}_\mathrm{AB}\hat{\sigma}_\mathrm{z}\otimes\hat{\sigma}_\mathrm{z}),
\end{equation}
with $\hat{\sigma}_\mathrm{z}=|0\rangle\langle0|-|1\rangle\langle1|$, so that
$$\hat{\sigma}_\mathrm{z}\otimes\hat{\sigma}_\mathrm{z}=|00\rangle\langle00|-|10\rangle\langle10|-|01\rangle\langle01|+|11\rangle\langle11|,$$
and where $\textrm{Tr}$ is the trace, defined as the sum of the diagonal matrix elements in any orthonormal basis, which in our case is $\textrm{Tr}\:\hat{O}=\sum_{i,j=0}^1\langle ij|\hat{O}|ij\rangle$, where $\hat{O}$ is an operator in $\mathcal{H}_{12}$.

By linearity of the trace, we can write
$$E_\mathrm{AB}=P_{00}-P_{10}-P_{01}+P_{11},$$
with
$$P_{kl}=\textrm{Tr}(\hat{\rho}_\mathrm{AB}|kl\rangle\langle kl|),$$
where $k$ and $l$ are 0 or 1.

The $P_{kl}$ are the joint probabilities, and we can now express them explicitly, using Eq.~(\ref{mixture}), as
$$P_{kl}=\sum_{i,j=0}^1\int_\Omega\rho(\lambda)d\lambda \: \langle ij|\Lambda_\mathrm{AB}\rangle\langle\Lambda_\mathrm{AB}|kl\rangle\langle kl|ij\rangle,$$
which, since $\{|00\rangle,|10\rangle,|01\rangle,|11\rangle\}$ forms an orthonormal basis in $\mathcal{H}_{12}$, simplifies as
\begin{equation}\label{Pkl}
    P_{kl}=\int_\Omega\rho(\lambda)d\lambda \:\big|\langle kl|\Lambda_\mathrm{AB}\rangle\big|^2.
\end{equation}

Note that, owing to the separability of $|\Lambda_\mathrm{AB}\rangle$ as expressed by Eq. (\ref{lambdaAB}), we can factorize the integrand in the above joint probability into a product, that is,
\begin{equation}\label{factorPkl}
    P_{kl}=\int_\Omega\rho(\lambda)d\lambda \:\big|\langle k|\lambda+\theta_\mathrm{A}\rangle \langle l|\lambda+\frac{\pi}{2}+\theta_\mathrm{B}\rangle\big|^2.
\end{equation}
With $P_i(\lambda,\theta)=\big|\langle i|\lambda+\theta\rangle\big|^2$ being the probability to get a photon in a channel $i$ for a local state $|\lambda\rangle$ and a measurement direction $\theta$ we can rewrite it as
\begin{equation}\label{jointprobexp}
P_{kl}=\int_\Omega\rho(\lambda)d\lambda \; P_k(\lambda,\theta_\mathrm{A})\; P_l(\lambda+\frac{\pi}{2},\theta_\mathrm{B}),
\end{equation}
which can be simply interpreted as the integral on all possible $\lambda$ of the product of the probability to detect a photon in channel $k$ on Alice's side for a local state $|\lambda\rangle$ by the probability to get a photon in channel $l$ on Bob's side for a local state $|\lambda+\frac{\pi}{2}\rangle$.

Assuming $\rho(\lambda)$ is a uniform distribution on the interval $[0,2\pi[$, we can then express these integrals explicitly as
\begin{equation}\label{jointprob}
    \begin{aligned}
P_{00}&=\frac{1}{2\pi}\int_0^{2\pi}d\lambda \cos^2(\lambda+\theta_\mathrm{A})\cos^2(\lambda+\frac{\pi}{2}+\theta_\mathrm{B}),\\
P_{10}&=\frac{1}{2\pi}\int_0^{2\pi}d\lambda \sin^2(\lambda+\theta_\mathrm{A})\cos^2(\lambda+\frac{\pi}{2}+\theta_\mathrm{B}),\\
P_{01}&=\frac{1}{2\pi}\int_0^{2\pi}d\lambda \cos^2(\lambda+\theta_\mathrm{A})\sin^2(\lambda+\frac{\pi}{2}+\theta_\mathrm{B}),\\
P_{11}&=\frac{1}{2\pi}\int_0^{2\pi}d\lambda \sin^2(\lambda+\theta_\mathrm{A})\sin^2(\lambda+\frac{\pi}{2}+\theta_\mathrm{B}),
    \end{aligned}
\end{equation}
which leads to
\begin{equation}\label{jointp}
    \begin{aligned}
    P_{00}&=P_{11}=\frac{1}{8} (2 - \cos 2(\theta_\mathrm{A} -\theta_\mathrm{B})),\\
    P_{10}&=P_{01}=\frac{1}{8} (2 + \cos 2(\theta_\mathrm{A} -\theta_\mathrm{B})),
    \end{aligned}
\end{equation}
and finally, we get
\begin{equation}\label{corrth}
    E_\mathrm{AB}=-\frac{1}{2}\cos2(\theta_\mathrm{A}-\theta_\mathrm{B}).
\end{equation}

This result only differs from the prediction for the singlet state by its visibility of 1/2 instead of 1, but with a maximum $S=\sqrt{2}$ it is clearly below 2 and therefore insufficient as an attack on Ekert protocol.

\section{Two-photon absorption attack}
In the two-photon absorption attack, Eve sends also pairs of pulses described by a mixture of separable states (\ref{mixture}) to Alice and Bob, but this time each pulse contains two photons instead of one. The photons inside a pulse share the same state: $|\lambda\rangle$ for the two photons sent to Alice; and $|\lambda+\frac{\pi}{2}\rangle$ for the two photons sent to Bob. We assume that the photons inside a pulse are independent: they follow the rules of Quantum Mechanics as prescribed by their quantum state independently of what the other photon is doing.

Now, the crucial difference is that these photons are chosen by Eve with a lower frequency than in the single-photon absorption case discussed above, so that the energy of a single photon is insufficient to trigger a click. Eve chooses the frequency of the photons such that the only way to get a click is through a two-photon absorption. We assume for simplicity that whenever two photons hit the same detector simultaneously, the probability that they trigger a click in the detector is 1. As we will see, this feature alone is (surprisingly) enough to lead to a clear violation of Bell inequalities.

So, a click can happen in a specific channel only when the two photons inside the same pulse choose to go to that same channel. If they choose different channels, no two-photon absorption can occur because there is only one photon in each channel. So, on each side, the three possibilities are:
\begin{itemize}
  \item Both photons go to channel 0 $\rightarrow$ click in channel 0 through a two-photon absorption,
  \item Both photons go to channel 1 $\rightarrow$ click in channel 1 through a two-photon absorption,
  \item One photon goes to channel 0, the other goes to channel 1 $\rightarrow$ no click in either channel.
\end{itemize}
Note that this third possibility brings us in the realm of the detection loophole \cite{Pearle,Larsson98,Gisin99,Adenier08,Adenier09}, which is the essential reason for the appearance of the violation of Bell inequalities that Alice and Bob are going to obtain. It should however be stressed that the non detections come from the frequency threshold in the photoelectric effect alone; a feature that is relevant in all detectors based on this effect, regardless of their quantum efficiency. Eve is therefore working in a fully Quantum Mechanical framework, without assuming anything about the detectors other than the existence of two-photon absorption processes at certain frequencies, and without assuming the existence of any hidden-variables.

We want to compute the probabilities $P_{kl}^{(2)}$ that Alice and Bob get a coincidence click respectively in channels $k$ and $l$. Obviously, the third case we considered above, that is, when a pulse does not produce any click on either side, is not going lead to any coincidence. We are therefore only concerned with the first two cases; those that can produce click.

Assuming independence between the photons, for a local state $|\lambda\rangle$ and a measurement angle $\theta$, on either side the probability that both photons from a same pulse end up in the same channel $i$ is simply the square of the probability $P_i(\lambda,\theta)$ to see one such photon going to this channel $i$, that is: $P_i^2(\lambda,\theta)$

So, similarly to what we had in the single photon case in Eq.(\ref{jointprobexp}), the probability $P_{kl}^{(2)}$ to get a click in channel $k$ for Alice and in channel $l$ for Bob in a two-photon absorption process is therefore the integral over all possible state of the product of the probability $P_k^2(\lambda,\theta_\mathrm{A})$ for Alice to get a click in channel $k$ by the probability $ P_l^2(\lambda+\frac{\pi}{2},\theta_\mathrm{B})$ for Bob to get a click in channel $l$:
\begin{equation}\label{jointprobexptwo}
P_{kl}^{(2)}=\int_\Omega\rho(\lambda)d\lambda \; P_k^2(\lambda,\theta_\mathrm{A})\; P_l^2(\lambda+\frac{\pi}{2},\theta_\mathrm{B}).
\end{equation}

For a rotationally invariant source, $\lambda$ is uniformly distributed on the interval $[0,2\pi[$, which leads to:
\begin{equation}\label{jointprobtp}
    \begin{aligned}
P_{00}^{(2)}&=\frac{1}{2\pi}\int_0^{2\pi}d\lambda \: \cos^4(\lambda+\theta_\mathrm{A})\cos^4(\lambda+\frac{\pi}{2}+\theta_\mathrm{B}),\\
P_{10}^{(2)}&=\frac{1}{2\pi}\int_0^{2\pi}d\lambda \: \sin^4(\lambda+\theta_\mathrm{A})\cos^4(\lambda+\frac{\pi}{2}+\theta_\mathrm{B}),\\
P_{01}^{(2)}&=\frac{1}{2\pi}\int_0^{2\pi}d\lambda \: \cos^4(\lambda+\theta_\mathrm{A})\sin^4(\lambda+\frac{\pi}{2}+\theta_\mathrm{B}),\\
P_{11}^{(2)}&=\frac{1}{2\pi}\int_0^{2\pi}d\lambda \: \sin^4(\lambda+\theta_\mathrm{A})\sin^4(\lambda+\frac{\pi}{2}+\theta_\mathrm{B}),
    \end{aligned}
\end{equation}
and we obtain:
\begin{equation}
    \begin{aligned}
P_{00}^{(2)}&=P_{11}^{(2)}=\frac{1}{128} (18 - 16 \cos 2(\theta_\mathrm{A} - \theta_\mathrm{B}) + \cos 4 (\theta_\mathrm{A} - \theta_\mathrm{B}),\\
P_{01}^{(2)}&=P_{10}^{(2)}=\frac{1}{128} (18 + 16 \cos 2(\theta_\mathrm{A} - \theta_\mathrm{B}) + \cos 4 (\theta_\mathrm{A} - \theta_\mathrm{B}).
    \end{aligned}
\end{equation}

Note that these four probabilities no longer add up to 1, because of the cases involving a non detection on either side or both, which are discarded by Alice and Bob. So, just like in a standard optical EPR experiment, the correlation function has to be normalized by the sum $\sum_{k,l}P_{kl}^{(2)}$:
\begin{equation}
E^{(2)}_\mathrm{AB}=
\frac{P_{00}^{(2)}-P_{10}^{(2)}-P_{01}^{(2)}+P_{11}^{(2)}}
{P_{00}^{(2)}+P_{10}^{(2)}+P_{01}^{(2)}+P_{11}^{(2)}},
\end{equation}
and that is explicitly:
$$E^{(2)}_\mathrm{AB}=-\frac{16 \cos 2(\theta_\mathrm{A} - \theta_\mathrm{B})}{18+\cos 4(\theta_\mathrm{A} - \theta_\mathrm{B})},$$
which lead to a violation of Bell inequalities for $\theta_\mathrm{A}=\{0,\frac{\pi}{4}\}$ and $\theta_\mathrm{B}=\{-\frac{\pi}{8},\frac{\pi}{8}\}$ of
$$S^{(2)}=\frac{16}{18}\;2\sqrt{2}\approx 2.51,$$
which is clearly above 2.

Note that there is slight dependence on $\cos 4(\theta_\mathrm{A} - \theta_\mathrm{B})$ that makes this correlation function differ from the $-V\cos 2(\theta_\mathrm{A} - \theta_\mathrm{B})$ expected for an entangled state (where $V$ is the visibility of the correlation in a real experiment), but the difference is quite small and would arguably be quite difficult to spot for Alice and Bob even if they were deciding to perform a full scan of the correlation.

The explanation for this seemingly unlikely violation of Bell inequalities with a mixture of separable states is that the sampling is unfair \cite{Adenier07,Adenier08,Adenier09}: the sample of pulses that get detected do not represent fairly the pulses that were emitted. In the case presented in the first section, where each pulse contained one photon only, the probabilities to get a click in either channel were adding up to one, that is, $P_0(\lambda,\theta)+P_1(\lambda,\theta)=1$, and were therefore treating all the pulses fairly. In the two-photon absorption case however, the probability on each side to get a click in either detector is
\begin{equation}\label{pclic}
\begin{aligned}
    P^{(2)}_\mathrm{clic}&=P_0^2(\lambda,\theta)+P_1^2(\lambda,\theta)\\
    &=\frac{1}{2}(1+\cos^2(2(\lambda+\theta)),
\end{aligned}
\end{equation}
which has a clear dependence on the state of the pulse $\lambda$, and is therefore not treating all the pulses fairly.

\section{Multiple-photon absorption attack}
A similar demonstration in the case of a three-photon absorption, with Eve sending three photons per pulse, leads to
\begin{equation}\label{jointprobthree}
P_{kl}^{(3)}=\int_\Omega\rho(\lambda)d\lambda \; P_k^3(\lambda,\theta_\mathrm{A})\; P_l^3(\lambda+\frac{\pi}{2},\theta_\mathrm{B})
\end{equation}
which exhibits a violation of Bell inequalities as high as $S^{(3)}\approx 3.17$. Quite generally, a multi-photon absorption process can lead to a violation of Bell inequalities as large as desired within the algebraic limit of 4, the only limit being the order of the multiple-photon absorptions that Eve can drive in Alice's and Bob's detectors.

Note that if Eve sends pulses meant to drive on one side a two-photon absorption and on the other a three-photon absorption (a feature that Eve could achieve by alternatively sending photons of different frequency to Alice and Bob), the relevant probabilities for the coincidences are of the form
\begin{equation}\label{jointprobexpthree}
P_{kl}^{(2,3)}=\int_\Omega\rho(\lambda)d\lambda \; P_k^2(\lambda,\theta_\mathrm{A})\; P_l^3(\lambda+\frac{\pi}{2},\theta_\mathrm{B})
\end{equation}
which leads to a correlation
$$E^{(2,3)}_\mathrm{AB}=-\frac{10 \cos 2(\theta_\mathrm{A} - \theta_\mathrm{B})}{10+\cos 4(\theta_\mathrm{A} - \theta_\mathrm{B})}$$
and a violation of Bell inequalities for $\theta_\mathrm{A}=\{0,\frac{\pi}{4}\}$ and $\theta_\mathrm{B}=\{-\frac{\pi}{8},\frac{\pi}{8}\}$ of exactly $$S^{(2,3)}=2\sqrt{2}.$$

\section{Discussion and possible countermeasures}
The violation of Bell inequalities observed by Alice and Bob is only apparent. If the non detected pulses were taken into account by Alice and Bob to compute the correlation, they would \emph{not} violate any Bell inequality. Eve's attack is nevertheless relevant because discarding the pairs for which no detection was recorded on either side and normalizing by the sum of coincidences is precisely how a violation of Bell inequality is observed in actual implementations of Ekert protocol with photons. Alice and Bob would therefore not be able to distinguish a genuine entangled state from this attack by simply observing a violation of Bell inequalities.

An unwanted feature of these attacks that could betray Eve's presence is that the sum of coincidences depends on the measurement settings $\theta_\mathrm{A}$ and $\theta_\mathrm{B}$. The stronger the violation of Bell inequalities, the stronger the visibility of the sum of coincidences. For instance, its visibility is about 0.06 in the two-photon absorption case ($S^{(2)}\approx 2.51$), and it is 0.10 in the mixed case with two-photon absorption on one side and three-photon on the other side ($S^{(2,3)}=2\sqrt{2}.$).

However, Eve can remove this unwanted effect entirely by driving different detection patterns for Alice and Bob, in a similar way to what was done by Larsson \cite{Larsson98} and Gisin \cite{Gisin99} in their hidden-variable models. The simplest method would be to alternatively drive a single photon absorption on one side, and a multiple-photon absorption on the other side. The sampling is then always fair on the side driven to a single photon absorption, and the total number of coincidences becomes independent of the measurement angles. It is nevertheless in Alice's and Bob's interest to closely monitor the sum of coincidence for any such angle dependence, because if it does not guarantee in principle against this attack, it makes Eve's task more difficult by forcing her to use higher order multiple-photon absorptions to achieve the same violation of Bell inequalities.

An obvious countermeasure against this multiple-photon absorption attack would be to use high-efficient detectors with low noise, so that undetected pulses would betray Eve's attack, but this is neither easy nor practical in an Ekert protocol with photons. Another possibility would be to thoroughly check and guarantee that no multiple-photon absorption can be dominant at any frequency in the detectors that are used by Alice and Bob. One way to limit this possibility would be to use highly selective frequency filters so that only a known range of restricted frequencies can reach the detectors, but it has the drawback of reducing all the more so the global efficiency of the channels. Last but not least, Alice and Bob could use our fair sampling test \cite{Adenier10}, as it does not introduce any loss and can be performed locally and unilaterally on either side during the production of the key.

\begin{acknowledgments}
Irina Basieva is supported by a grant from the Swedish Institute.
\end{acknowledgments}

\end{document}